\documentclass[aip,jcp,reprint]{revtex4-1}
\pdfoutput=1

\usepackage{graphicx}
\usepackage{amsmath}
\usepackage{amssymb}
\usepackage{multirow}
\usepackage{float}
\usepackage{hyperref}
\usepackage{color}

\newcommand{\Angstrom}{\ensuremath{\mathring{\textnormal{A}}}}
 \newcommand{\sub}[1]{\ensuremath{_{\textrm{#1}}}} \newcommand{\super}[1]{\ensuremath{^{\textrm{#1}}}} 
\renewcommand{\vec}[1]{\mathbf{#1}}

\newcommand{\JCAP}{Joint Center for Artificial Photosynthesis, California Institute of Technology, Pasadena, CA 91125}

\begin{document}

\title{Grand canonical electronic density-functional theory:\\algorithms and applications to electrochemistry}

\author{Ravishankar Sundararaman}\email{sundar@rpi.edu}
\affiliation{Department of Materials Science and Engineering, Rensselaer Polytechnic Institute, Troy, NY 12180}\affiliation{\JCAP}
\author{William A. Goddard III}\email{wag@wag.caltech.edu}
\affiliation{Materials and Process Simulation Center, California Institute of Technology, Pasadena, CA 91125}\affiliation{\JCAP}
\author{Tomas A. Arias}\email{taa2@cornell.edu}
\affiliation{Laboratory of Atomic and Solid State Physics, Cornell University, Ithaca, NY 14853}
\date{\today}

\begin{abstract}
First-principles calculations combining density-functional theory and continuum solvation
models enable realistic theoretical modeling and design of electrochemical systems. 
When a reaction proceeds in such systems, the number of electrons in the portion of the system treated
quantum mechanically changes continuously, with a balancing charge appearing in the continuum electrolyte.
A grand-canonical ensemble of electrons at a chemical potential set by the electrode potential
is therefore the ideal description of such systems that directly mimics the experimental condition.
We present two distinct algorithms, a self-consistent field method (GC-SCF) and a direct variational
free energy minimization method using auxiliary Hamiltonians (GC-AuxH), to solve the Kohn-Sham equations
of electronic density-functional theory directly in the grand canonical ensemble at fixed potential.
Both methods substantially improve performance compared to a sequence of conventional
fixed-number calculations targeting the desired potential, with the GC-AuxH method additionally
exhibiting reliable and smooth exponential convergence of the grand free energy.
Finally, we apply grand-canonical DFT to the under-potential deposition of copper on
platinum from chloride-containing electrolytes and show that chloride desorption,
not partial copper monolayer formation, is responsible for the second voltammetric peak.
\end{abstract}

\maketitle

\makeatletter{}Density-functional theory (DFT) enables theoretical elucidation
of reaction mechanisms at complex catalyst surfaces, making it now possible
to design efficient heterogeneous catalysts for various industrial applications
from first principles, for example for high-temperature gas-phase transformation
of hydrocarbons to a variety of valuable chemical products.\cite{Ammoxidation,Anhydride}
The extension of this predictive power to electrocatalysis
would be highly valuable for an even broader class of technological problems,
including a cornerstone of future technology for renewable energy:
converting solar energy to chemical fuels by electrochemical
water splitting and carbon dioxide reduction.\cite{SolarWaterSplitting}
Accurately describing electrochemical phenomena, however, presents two additional challenges.

First, the electrolyte, typically consisting of ions in a liquid solvent,
strongly affects the energetics of structures and reactions at the interface.
Treating liquids directly in DFT requires expensive molecular dynamics to
sample the thermodynamic phase space of atomic configurations.
Historically, a number of continuum solvation models that empirically capture liquid effects
have enabled theoretical design of liquid-phase catalysts.\cite{PCM-SMD,PCM-Review}
More recently, empirical solvation models suitable for solid-liquid interfaces,\cite{PCM-SCCS,NonlinearPCM,VASPsol}
joint density-functional theory (JDFT) for efficiently treating liquids with atomic-scale structure,\cite{JDFT}
and minimally-empirical solvation models derived from JDFT,\cite{SaLSA,CANDLE} have made
great strides towards reliable yet efficient treatment of electrochemical systems.

Second, electrons can flow in and out of the electrode as electrochemical reactions proceed.
Changes in electronic charge of electrode surfaces and adsorbates can be especially important because
the electrolyte stabilizes charged configurations with a counter charge from the ionic response.
For example, reduction of formic acid on platinum at experimentally relevant potentials
is dominated by formate ions rather than neutral molecules at the surface.\cite{FormateOxidation}
Proton adsorption on stepped and polycrystalline surfaces involves displacing
oxidatively-adsorbed water at relevant potentials, resulting in non-integer charge transfers
and an anomalous pH dependence deviating from the Nernst equation.\cite{AnomalousPH}

Accounting for the electrolyte response using our solvation models,\cite{CANDLE,VASPsol}
and adjusting the electron number to match experimentally relevant electrode potentials,
realistic predictions of electrochemical reaction mechanisms have now become possible.\cite{UadjustLoop}
In particular, application of this methodology to the reduction of CO on Cu(111)
predicts onset potentials for methane and ethene formation with 0.05~V accuracy
in comparison to experiment, for a wide range of pH varying from 1 to 12.\cite{COreductionCu111}
However, conventional DFT software and algorithms are optimized for solving the
quantum-mechanical problem at fixed electron number, requiring extra work (both
manual and computational) to calculate properties for a specified electrode potential.

Electric potentials and fields play an important role in fields besides than electrochemistry.
Density-functional theory approaches accounting for electric potential have been developed
in special cases for field emission from metal surfaces using a jellium
model,\cite{FieldEmission} and for calculating capacitance in metal-insulator-metal\cite{FiniteFieldMIM,FiniteBiasOSA} and carbon nanotube systems.\cite{NanotubeCapacitance}
Calculating non-equilibrium transport of electrons in nanoscale systems also requires
accounting for potential difference between reservoirs in a DFT calculation.\cite{DFT-NEGF}
First-principles molecular dynamics approaches have been developed to emulate
fixed potential using fluctuating numbers of electrons between time steps.\cite{FixedPotentialAIMD}
However, in all these cases, each involved self-consistent DFT calculation contains a fixed
number of electrons and is carried out using a conventional canonical-ensemble algorithm.

This paper introduces algorithms for grand canonical DFT,
where electron number adjusts automatically to target a specified
electron chemical potential (related to electrode potential), thereby enabling
efficient and intuitive first-principles treatment of electrochemical phenomena.
Section~\ref{sec:Theory} summarizes the theoretical background of first-principles electrochemistry
using JDFT and continuum solvation models, and sets up the fundamental basis of grand-canonical DFT.
Then, section~\ref{sec:Algorithms} introduces the modifications necessary to
make two distinct classes of DFT algorithms, the self-consistent field (GC-SCF) method
and the variational free energy minimization using auxiliary Hamiltonians (GC-AuxH),
directly converge the grand free energy of electrons at fixed potential.
Sections~\ref{sec:ResultsSCF} and \ref{sec:ResultsAux} establish the algorithm parameter(s)
that optimize the iterative convergence of the GC-SCF and GC-AuxH methods respectively,
while section~\ref{sec:AlgoCompare} compares the performance of these algorithms
for a number of prototypical electrochemical systems.
Finally, section~\ref{sec:UPD} demonstrates the utility of grand canonical DFT
by solving an electrochemical mystery: the identity of the second voltammetric peak in the
under-potential deposition (UPD) of copper on platinum in chloride-containing electrolytes.
 
\makeatletter{}\section{Theory}\label{sec:Theory}

\subsection{Background: electronic density functional theory}

The exact Helmholtz free energy $A$ of a system of interacting electrons
in an external potential $V(\vec{r})$ at a finite temperature $T$
satisfies the Hohenberg-Kohn-Mermin variational theorem\cite{HK-DFT,ThermalDFT-Mermin}
\begin{equation}
A = \min_{n(\vec{r})} \left( A\sub{HKM}^T[n(\vec{r})] + \int d\vec{r} V(\vec{r}) n(\vec{r}) \right),
\label{eqn:DFT-HKM}
\end{equation}
where $A\sub{HKM}^T$ is a universal functional that depends
only on the electron density $n(\vec{r})$ (and temperature),
and not on the external potential.
However, constructing approximations for this unknown universal functional
that accurately capture the energies and geometries of chemical bonds
in terms of the density alone is extremely challenging,
partly because the quantum mechanics of the electrons
is completely implicit in $A\sub{HKM}[n]$
(dropping the $T$ labels here onward for notational simplicity;
all the functionals below depend on temperature).

Most practical approximations in electronic density-functional theory
follow the Kohn-Sham approach\cite{KS-DFT} that includes the exact free energy of
a non-interacting system of electrons with the same density $n(\vec{r})$.
The universal functional is typically split as
\begin{equation}
A\sub{HKM}[n] = A\sub{ni}[n]
	+ \underbrace{\int d\vec{r} \int d\vec{r}' \frac{n(\vec{r}) n(\vec{r}')}{2|\vec{r}-\vec{r}'|}}_{E_H[n]}
	+ E\sub{XC}[n],
\label{eqn:DFT-KS}
\end{equation}
where $A\sub{ni}[n]$ is the non-interacting free energy (which we describe in detail below),
the second `Hartree' term $E_H[n]$ is the mean-field Coulomb interaction between electrons
(using atomic units $e,m_e,\hbar,k_B=1$ throughout), and the final `exchange-correlation'
term $E\sub{XC}[n]$ captures the remainder which is not known and must hence be approximated.
The exact free energy of non-interacting electrons is
\begin{equation}
A\sub{ni}[n] = \min_{
	\mbox{\scriptsize$\begin{array}{c}
	\{\psi_i(\vec{r}),f_i\}\\
	\rightarrow n(\vec{r})
	\end{array}$}
}
\sum_i \bigg(
	\underbrace{\frac{f_i}{2} \int d\vec{r} |\nabla\psi_i(\vec{r})|^2}\sub{Kinetic}\\
	- T\underbrace{S(f_i)}\sub{Entropy}
\bigg),
\label{eqn:DFT-ni}
\end{equation}
which includes the kinetic energy and entropy contributions of a set of orthonormal
single-particle orbitals $\psi_i(\vec{r})$ with occupation factors $f_i \in [0,1]$.
Above, the single-particle entropy function is $S(f) = -f \log f - (1-f)\log(1-f)$.
Note that we let the orbital index $i$ include spin degrees of freedom as well,
and therefore do not introduce factors of 2 for spin degeneracy.
For notational convenience, we also let $i$ include Bloch wave-vectors in the
Brillouin zone (in addition to spin and band indices) for periodic systems.

The minimization in (\ref{eqn:DFT-ni}) is constrained such that the density of the non-interacting system
\begin{equation}
\sum_i f_i |\psi_i(\vec{r})|^2 = n(\vec{r}),
\label{eqn:DensityConstraint}
\end{equation}
the density of the real interacting system.
Performing this minimization with Lagrange multipliers $V\sub{KS}(\vec{r})$ (the Kohn-Sham potential)
to enforce the density constraint and $\varepsilon_i$ (Kohn-Sham eigenvalues) to enforce
orbital normalization constraints, results in a set of single-particle Schr\"odinger-like equations
\begin{equation}
-\frac{\nabla^2}{2} \psi_i(\vec{r}) + V\sub{KS}(\vec{r})\psi_i(\vec{r}) = \varepsilon_i \psi_i(\vec{r})
\label{eqn:SchrodingerKS}
\end{equation}
for stationarity with respect to $\psi_i(\vec{r})$,
and the Fermi occupation condition
\begin{equation}
f_i = \frac{1}{1+\exp\frac{\varepsilon_i-\mu}{T}}
\label{eqn:FermiOccupations}
\end{equation}
for stationarity with respect to $f_i$.
Here, the electron chemical potential $\mu$ appears as a Lagrange multiplier
to enforce the electron number constraint, and is chosen so that $\sum f_i = N$,
the number of electrons in the system.

Optimizing the total free energy functional (\ref{eqn:DFT-HKM})
with $A\sub{HKM}[n]$ implemented by (\ref{eqn:DFT-KS},\ref{eqn:DFT-ni})
then yields the stationarity condition with respect to electron density,
\begin{equation}
V\sub{KS}[n](\vec{r}) = V(\vec{r}) + \frac{\delta}{\delta n(\vec{r})}
	\left( E_H[n] + E\sub{XC}[n] \right).
\label{eqn:V_KS}
\end{equation}
Conventional density-functional theory calculations then amount to self-consistently
solving the Kohn-Sham equations (\ref{eqn:SchrodingerKS}) along with (\ref{eqn:V_KS}),
coupled via the electron density constraint (\ref{eqn:DensityConstraint}).

The Kohn-Sham potential is arbitrary up to an overall additive constant:
changing this constant introduces a rigid shift in the eigenvalues $\epsilon_i$
and the electron chemical potential $\mu$, but does not affect
the occupations $f_i$, electron density or free energy.
For finite systems (of any charge) and for neutral systems that are infinitely
periodic in one or two directions, this arbitrariness can be eliminated by
requiring that the potential vanishes infinitely far away from the system,
giving meaning to the absolute values of $\mu$ and $\epsilon_i$
as being referenced to `zero at infinity'.
However, for materials that are infinite in all three dimensions,
such as periodic crystalline solids, there is no analogous 
natural choice for the zero of potential, making the absolute
reference for $\mu$ and $\epsilon_i$ meaningless.

More importantly, for systems that are periodic in one, two or all three directions,
the net charge per unit cell must be zero, otherwise the energy per unit cell becomes infinite.
In terms of a finite system size $L$ and then taking the limit $L\rightarrow\infty$,
the energy per unit cell of a system with net charge per unit cell
diverges $\propto \ln L$ for one periodic direction, $\propto L$ for
two periodic directions and $\propto L^2$ for three periodic directions.
Therefore, the number of electrons per unit cell is physically constrained
to keep the unit cell neutral in systems with \emph{any} periodicity,
and only finite systems (like molecules and ions) have
number of electrons as a degree of freedom.

\subsection{Electrochemistry with joint density-functional theory}\label{sec:TheoryJDFT}

For describing electrochemical systems and electrocatalytic mechanisms,
we are typically interested in adsorbed species in a solid-electrolyte
interface which exchange electrons with the solid (electrode).
In these systems, the solid surface, which we would describe
in a density-functional calculation as a slab periodic in two directions,
does have a net charge per unit cell that depends on the electrode potential.
In contrast to the discussion at the end of the previous section,
this is physically possible (i.e. has a finite energy)
because the electrolyte contains mobile ions that respond by
locally increasing the concentration of ions of opposite charge
near the surface, thereby neutralizing the unit cell.
(The charge per unit areas of the electrode
and electrolyte are equal and opposite.)

Next to an electrolyte, the charge of a partially-periodic system
(one-dimensional `wires' or two-dimensional `slabs') is no longer
constrained, allowing the number of electrons to vary.
Now, the absolute reference for the electron chemical potential $\mu$ does
become physically meaningful, and $\mu$ now corresponds to the electrode
potential that controls the number of electrons in the electrode.

However, treating electrochemical systems using electronic density-functional
theory alone is extremely challenging for a variety of reasons.
First, treating liquids requires a statistical average over a large number of
atomic configurations to integrate over thermodynamic phase space.
For this, techniques such as molecular dynamics typically require
calculations of at least $10^4 - 10^5$ atomic configurations
(instead of just one for a solid).
Second, such calculations require a large number of liquid molecules
to minimize finite size errors in the molecular dynamics.
For example, for electrolytes with a realistic ionic concentration of 0.1~M,
there is \emph{on average} one ion for a few hundred solvent molecules.
Making statistical errors in the ion number manageable in such calculations
therefore requires $>10^3$ solvent molecules with $>10^4$ electrons,
contrasted with a typical $10-100$ atoms with $100-10^3$ electrons
in the electrode slab + adsorbate of interest.
Combined, these factors make density-functional molecular dynamics
simulations of electrochemical systems prohibitively expensive
computationally, in additional to being difficult to set up and analyze.

A viable alternative to the above direct approach is to employ
joint density-functional theory\cite{JDFT} (JDFT), a variational theorem
akin to the Hohenberg-Kohn theorem that makes it possible to describe the free energy
of a solvated system in terms of the electron density $n(\vec{r})$ for the solute
and in terms of a set of nuclear densities $\{N_\alpha(\vec{r})\}$
(where $\alpha$ indexes nuclear species) of the solvent (or electrolyte).
Specifically, the equilibrium free energy of the combined solute and solvent systems minimizes
\begin{multline}
A = \min_{n(\vec{r}),\{N_\alpha(\vec{r})\}}
	\bigg( \tilde{A}\sub{JDFT}[n(\vec{r}),\{N_\alpha(\vec{r})\}] \\
	+ \int d\vec{r} V(\vec{r}) n(\vec{r})
	+ \sum_\alpha \int d\vec{r} V_\alpha(\vec{r}) N_\alpha(\vec{r}) \bigg),
\label{eqn:JDFT-theorem}
\end{multline}
where $V(\vec{r})$ is the external electron potential,
$V_\alpha(\vec{r})$ is the external potential on the liquid nuclei and
$A\sub{JDFT}$ is a universal functional independent of these external potentials.
Separating out the Hohenberg-Kohn electronic density functional $A\sub{HK}[n]$
for the solute, the total free energy is
\begin{equation}
\tilde{A}\sub{JDFT}[n,\{N_\alpha\}]
 = \underbrace{A\sub{HKM}[n]}\sub{electronic}
 + \underbrace{\tilde{A}\sub{diel}[n,\{N_\alpha\}]}\sub{solvation}.
\label{eqn:JDFT}
\end{equation}
In practice, the functional $A\sub{diel}$, much like $A\sub{HKM}$,
is unknown and needs to be approximated.
Importantly, the liquid is now described directly in terms of its average density
rather than individual configurations, and the expensive quantum-mechanical theory
of the electrons is restricted to the solute alone, thereby addressing both the sampling
and system-size problems that make density-functional molecular dynamics prohibitively expensive.

Typically, we are interested in a situation where the external potentials
on the liquid are zero, and the liquid only interacts with itself,
and with the electrons and nuclei of the solute.
In this situation, we can perform the optimization over liquid densities
and define the implicit functional $A\sub{JDFT}[n] = A\sub{HKM}[n] + A\sub{diel}[n]$
where we define $A_X[n]\equiv \min_{\{N_\alpha(\vec{r})\}} \tilde{A}_X[n(\vec{r}),
\{N_\alpha(\vec{r})\}]$ for $X$ = JDFT and $X$ = diel.
We will work with these reduced functionals below for simplicity.

The framework of joint density-functional theory encompasses an entire hierarchy of solvation theories.
Further separating the solvation term $A\sub{diel}$ into a classical density functional
for the liquid\cite{BondedTrimer,RigidCDFT,PolarizableCDFT} and an electron-liquid interaction functional,
unlocks the full potential of JDFT to describe atomic-scale structure in the liquid without statistical sampling.
Starting from `full-JDFT' and performing perturbation theory for linear-response
of the liquid results in the non-empirical SaLSA solvation model\cite{SaLSA}
which continues to capture the atomic-scale nonlocality in the liquid response
and introduces no fit parameters for the electric response of the solvent.

At the simplest end of the JDFT hierarchy, are continuum solvation models that neglect
the nonlocality of the liquid response and replace it by that of an empirically-determined
dielectric cavity (optionally with Debye screening due to electrolytes).
This includes our recent CANDLE solvation model\cite{CANDLE}
that builds on the stability of SaLSA for highly polar systems,
and earlier solvation models suitable for molecules and less-polar systems
such as our GLSSA13 model\cite{NonlinearPCM} (or its equivalent, VASPsol\cite{VASPsol})
and the comparable Self-Consistent Continuum Solvation (SCCS) model.\cite{PCM-SCCS,PCM-SCCS-charged}
Even the traditional quantum-chemistry finite-system solvation models such as the PCM series\cite{PCM-Review}
and the SMx series\cite{PCM-SMD} can be mapped on to this class of solvation models.

For simplicity, we will work here with this simplest class of continuum solvation models.
The theoretical considerations and algorithms in this work focus primarily
the electronic density-functional theory component, are largely agnostic to
the internals of the solvation model and therefore straightforwardly generalize
up the JDFT hierarchy to the more detailed and complex solvation theories.
Essentially, all the simple electron-density-based continuum solvation
models\cite{CANDLE,NonlinearPCM,VASPsol,PCM-SCCS} can be summarized
abstractly as\cite{NonlinearPCM,SaLSA}
\begin{equation}
A\sub{diel}[n(\vec{r})] = \int d\vec{r} \rho\sub{el}(\vec{r})
\frac{(\hat{K}^{-1}-\hat{\chi})^{-1}-\hat{K}}{2}
\rho\sub{el}(\vec{r}) + A\sub{cav}[s].
\label{eqn:AdielPCM}
\end{equation}
Here, the first term is the electrostatic solute-solvent interaction energy
given by the difference between the solvent-screened Coulomb interaction
and the bare Coulomb interaction $\hat{K}$ of the total solute charge
density, $\rho\sub{el}(\vec{r}) = n(\vec{r}) + \rho\sub{nuc}(\vec{r})$,
where $\rho\sub{nuc}(\vec{r})$ is the solute nuclear charge density.
The screened Coulomb interaction term above is expressed in terms of the
bare Coulomb interaction $\hat{K}$ and the solvent susceptibility
operator $\hat{\chi}$, defined for the local-response models by
\begin{equation}
\hat{\chi}\cdot\phi(\vec{r}) \equiv 
	\underbrace{\nabla\cdot\left(\frac{\epsilon_b-1}{4\pi}s(\vec{r})\nabla\phi(\vec{r})\right)}\sub{Dielectric}
	-\underbrace{\frac{\kappa^2}{4\pi}s(\vec{r}) \phi(\vec{r})}\sub{Ionic}.
\label{eqn:PCMchi}
\end{equation}
Here, $\epsilon_b$ is the bulk dielectric constant of the solvent
and $\kappa = \sqrt{4\pi \sum_i N_i Z_i^2 / T}$ is the inverse Debye
screening length in vacuum of a set of ionic species of charge $Z_i$
with concentrations $N_i$ (the net inverse Debye screening length
in the presence of the background dielectric is $\kappa/\sqrt{\epsilon_b}$).
The cavity shape function $s(\vec{r})$ modulates the dielectric and ion
response, and varies from zero in the solute region (no solvent present)
to unity in the solvent at full bulk density.
Finally, the second term of (\ref{eqn:AdielPCM}), $A\sub{cav}$,
empirically captures effects beyond mean-field electrostatics,
such as the free energy of cavity formation in the liquid and
dispersion interactions between the solute and solvent.\cite{CavityWDA}
Different solvation models at this level of the hierarchy only differ in
the details of how $s(\vec{r})$ is determined from the electron density
(or even from atomic positions and fit radii for the PCM\cite{PCM-Review}
and SMx\cite{PCM-SMD} solvation models), and in the details of $A\sub{cav}$.

In practice, evaluating the first term of (\ref{eqn:AdielPCM}) requires the calculation
of $\phi(\vec{r}) = (\hat{K}^{-1}-\hat{\chi})^{-1} \rho\sub{el}(\vec{r})$,
which is the net electrostatic potential including screening by the solvent (and electrolyte).
The second part of the first term in (\ref{eqn:AdielPCM}) represents the electrostatic interactions between
electrons and the nuclei in vacuum which cancels corresponding terms in the vacuum DFT functional
(specifically the Hartree, Ewald and long-range part of the local pseudopotential terms),
so that all long-range terms that contribute to the net free energy
are present in the first term of (\ref{eqn:AdielPCM}).
Finally, substituting (\ref{eqn:PCMchi}) and $\hat{K}^{-1}=-\nabla^2/(4\pi)$
into the definition of $\phi(\vec{r})$ results in the linearized
Poisson-Boltzmann (modified Helmholtz) equation
\begin{equation}
-\nabla\cdot\left(\epsilon_b s(\vec{r})\nabla\phi(\vec{r})\right) 
+\kappa^2 s(\vec{r})\phi(\vec{r}) = 4\pi\rho\sub{el}(\vec{r}).
\label{eqn:LinearPBE}
\end{equation}

Without ionic screening (second term of (\ref{eqn:LinearPBE}) above),
the absolute reference for $\phi(\vec{r})$ is undetermined and does
not contribute to the bound charge charge induced in the liquid,
$\hat{\chi}\phi(\vec{r})$ (given by the first term of (\ref{eqn:PCMchi}) alone),
because $\nabla(\mathrm{constant})=0$.
However, with ionic screening, a constant shift of $\phi(\vec{r})$
does affect the second terms of (\ref{eqn:PCMchi}) and (\ref{eqn:LinearPBE}),
producing a charge response in the liquid and making the absolute
reference of the electrostatic potential $\phi(\vec{r})$ meaningful.
In particular, integrating (\ref{eqn:LinearPBE}) over space yields
\begin{equation}
\int d\vec{r} \frac{\kappa^2}{4\pi} s(\vec{r})\phi(\vec{r}) = \int\rho\sub{el}(\vec{r}).
\end{equation}
From (\ref{eqn:PCMchi}), the left hand side is the negative of the total bound charge
in the liquid, while the right hand side is the total charge of the solute system.
Therefore, the continuum electrolyte automatically compensates for
any net charge in the solute and makes the complete system neutral.
As a side effect, the absolute reference of the electrostatic potential
is meaningful and automatically corresponds to `zero at infinity'.
This happens because the Greens function of (\ref{eqn:LinearPBE}) is
$\exp(-\kappa r/\sqrt{\epsilon_b}) / r$ in the bulk liquid with finite $\kappa$
(instead of $1/r$), causing $\phi(\vec{r})$ to exponentially decay
to zero outside the solute region (where $\rho\sub{el}=0$).

Ref.~\citenum{PCM-Kendra} gives a detailed version of the above discussion
including rigorous proofs of the absoluteness of the reference for $\phi(\vec{r})$,
and numerical details and algorithms for efficiently solving (\ref{eqn:LinearPBE})
with periodic boundary conditions in plane-wave basis DFT calculations.
The result that the electrolyte neutralizes charges in the solute
is true more generally, even for solvation models with
nonlinear\cite{NonlinearPCM} and/or nonlocal response.\cite{SaLSA}
However, as Ref.~\citenum{NonlinearPCM} describes in detail,
for the general nonlinear case, unit cell neutralization is not
automatic and must be imposed using a Lagrange multiplier constraint;
this Lagrange multiplier then fixes the absolute value
of $\phi(\vec{r})$ such that it is zero at infinity.

The key conclusion of this discussion is that continuum electrolytes present two advantages.
First, automatic charge neutralization implies we can perform
well-defined calculations with net charge per unit cell in the solute.
Second, meaningful absolute reference values for potentials implies that
the electron chemical potential $\mu$ (which determines the electron occupations
in (\ref{eqn:FermiOccupations})) is also referenced to zero at infinity.
Consequently, $\mu$ is related directly to the potential $U$ of
the electrode providing the electron reservoir in experiments.
This potential is typically referenced to the standard hydrogen electrode (SHE), so that
\begin{equation}
\mu = \mu\sub{SHE} - U,
\label{eqn:muCalibration}
\end{equation}
where $\mu\sub{SHE}$ is the absolute position of the standard hydrogen electrode
relative to the vacuum level (i.e. the zero at infinity reference).
Calculations can employ either the experimental estimate $\mu\sub{SHE}=-4.44$~eV
to relate the absolute and relative potential scales,\cite{SHE-IUPAC,SHE-comparison}
or a theoretical calibration based on the calculated and measured
potentials of zero charge of solvated metal surfaces.\cite{PCM-Kendra}
The latter approach has the advantage of minimizing
systematic errors in the solvation model since they cancel
between the calculations used for calibration and prediction.
For example, with the CANDLE solvation model we use below,
the calibrated $\mu\sub{SHE} = -4.66$~eV.\cite{CANDLE}

\subsection{Grand-canonical density-functional theory}\label{sec:TheoryFixedPotential}

Using joint density-functional theory of continuum solvation models
to treat electrolytes as described above, we can calculate the
Helmholtz free energy and electrochemical potential ($\mu$, and hence $U$)
of specific microscopic configurations of adsorbates on electrode surfaces
at a fixed number of electrons $N$ in the solute subsystem (electrode + adsorbates).
However, in electrochemical systems, $N$ is an artificial constraint because
electrons can freely exchange between the electrode and an external circuit.
Instead, experiments set the electrode potential $U$, and hence the electron
chemical potential $\mu$, and $N$ adjusts accordingly as a dependent variable.
Thermodynamically, this corresponds to switching the electrons
from the finite-temperature, fixed-number canonical ensemble to
the finite-temperature, fixed-potential grand-canonical ensemble.
Correspondingly, the relevant free energy minimized at equilibrium is the
grand free energy $\Phi = A - \mu N$, instead of the Helmholtz free energy $A$.

The most straightforward approach to fixed-potential DFT for electrochemistry
is to repeat conventional fixed-charge DFT calculations at
various electron numbers $N$ to reach a target chemical potential $\mu$.
Using a steepest descent / secant method to optimize $N$ results in
convergence of $\mu$ typically within 10 iterations.\cite{UadjustLoop}
However this approach is inefficient since it requires multiple DFT calculations
to calculate the grand free energy and charge of a single configuration,
typically taking 3x the time of a single fixed-charge calculation (not 10x because
subsequent calculations have a better starting point and converge quicker).\cite{UadjustLoop}

A more efficient approach would be to directly optimize
the Kohn-Sham functional in the grand-canonical ensemble.
For a solvated system, this corresponds to a small modification of the minimization problem
(\ref{eqn:JDFT-theorem}) with $A\sub{JDFT}$ given by (\ref{eqn:JDFT})
and with $A\sub{HKM}$ evaluated using the Kohn-Sham approach
(\ref{eqn:DFT-KS},\ref{eqn:DFT-ni}).
The only differences are that the Lagrange multiplier term
$-\mu (\sum_i f_i - N)$ that enforced the electron number constraint
is replaced by $-\mu \sum_i f_i$ which implements the Legendre transform of the
Helmholtz free energy to the grand free energy, and that the fixed-$N$ constraint is removed.
The electron occupation factors are still Fermi functions (\ref{eqn:FermiOccupations}),
but $\mu$ is specified as an input (instead of being adjusted to match a specified $N$).
Operationally, this amounts again to solving the Kohn-Sham eigenvalue problem
(\ref{eqn:SchrodingerKS}) self-consistently with
\begin{equation}
V\sub{KS}[n](\vec{r}) = V(\vec{r}) + \frac{\delta}{\delta n(\vec{r})}
	\left( E_H[n] + E\sub{XC}[n] + A\sub{diel}[n]\right),
\label{eqn:V_KS_sol}
\end{equation}
where there is now a single extra contribution to the potential due to the solvent (electrolyte),
and with the aforementioned changes to the Lagrange multipliers and constraints.

This conceptually simple modification, however, presents numerical challenges
to the algorithms commonly used for solving the Kohn-Sham problem,
such as the self-consistent field (SCF) method, which
solves the Kohn-Sham eigenvalue problem from an input density
(or Kohn-Sham potential), adjusting this density (or potential)
iteratively until self-consistency is achieved.
A common instability for metallic systems in the SCF method is `charge sloshing':
the electron density oscillates spatially between iterations instead of converging.
Switching to the grand-canonical ensemble can significantly exacerbate this problem:
electrons can now additionally slosh between the system and the electron reservoir.
Here, we present modifications to the SCF method and an alternate algorithm
that allow reliable and efficient convergence for \emph{grand-canonical} Kohn-Sham DFT.
 
\makeatletter{}\section{Algorithms}\label{sec:Algorithms}

\subsection{Self-Consistent Field method: Pulay mixing}\label{sec:AlgoSCF}

Given electron density $n^{(i)}\sub{in}(\vec{r})$, the Kohn-Sham equations
(\ref{eqn:SchrodingerKS}) with potential $V\sub{KS}$ given by (\ref{eqn:V_KS_sol})
define orbitals $\{\psi_j\}$ and eigenvalues $\{\varepsilon_j\}$.
At a given electron chemical potential $\mu$, these in turn define
the occupations $\{f_j\}$ given by (\ref{eqn:FermiOccupations})
and a new electron density $n^{(i)}\sub{out}(\vec{r})$
given by (\ref{eqn:DensityConstraint}).
The Self-Consistent Field (SCF) method attempts to find $n^{(i)}(\vec{r})$
such that $n^{(i)}\sub{out}(\vec{r}) = n^{(i)}\sub{in}(\vec{r})$.

There are two algorithmic ingredients to this method:
solution of the Kohn-Sham eigenvalue equations
and optimization of the electron density.
The eigenvalue equations remain unchanged between
conventional and fixed-potential calculations.
To solve, these we use the standard Davidson algorithm.\cite{Davidson}
The difficulty in fixed-potential calculations arise
in the electron density optimization, which we discuss below.

A robust and frequently-used algorithm for charge-density optimization
is Pulay mixing\cite{PulayMixing} with Kerker preconditioning.\cite{KerkerMixing}
Briefly, Pulay mixing assumes that the residual $R[n\sub{in}(\vec{r})]
\equiv n\sub{out}(\vec{r}) - n\sub{in}(\vec{r})$ is approximately linear
in the input electron density $n\sub{in}(\vec{r})$, and calculates
the optimum input electron density as a linear combination of previous iterations,
\begin{equation}
n\super{opt}\sub{in}(\vec{r}) = \sum_i \alpha_i n^{(i)}\sub{in}(\vec{r})
\end{equation}
Minimization of the norm of the corresponding residual
\begin{equation}
F(\{\alpha_i\}) = \sum_{ij} \alpha_i \alpha_j \int d\vec{r}
	R[n^{(i)}\sub{in}(\vec{r})] \hat{M} R[n^{(j)}\sub{in}(\vec{r})]
\end{equation}
with the constraint $\sum_i \alpha_i = 1$ yields a set of
linear equations determining the coefficients $\alpha_i$,
where $\hat{M}$ is the metric for defining the norm of the residual.
Finally, the next input density is obtained by mixing the
optimum input density with its corresponding output density
\begin{equation}
n^{(i+1)}\sub{in}(\vec{r}) = n\super{opt}\sub{in}(\vec{r})
	+ \hat{K} R[n\super{opt}\sub{in}(\vec{r})],
\end{equation}
where $\hat{K}$ is the Kerker preconditioning operator.

The metric $\hat{M}$ and preconditioner $\hat{K}$ serve to
balance the influence of different components of the
electron density to the optimization procedure.
These are usually defined in reciprocal space, expanding $n(\vec{r})
= \sum_{\vec{G}} \tilde{n}(\vec{G}) e^{i\vec{G}\cdot\vec{r}}$,
where $\vec{G}$ are reciprocal lattice vectors.
In reciprocal space, the Hartree potential in (\ref{eqn:V_KS_sol})
takes the form $\tilde{V}_H(\vec{G}) = \tilde{n}(\vec{G}) 4\pi/G^2$,
causing small $G$ (long-wavelength) variations of the electron density
to produce larger changes in the Kohn-Sham potential than large $G$ ones.
Uncompensated, this makes the electron density optimization unstable against
long-wavelength perturbations (the charge-sloshing instability).
The Kerker preconditioner
\begin{equation}
\tilde{K}(\vec{G}) = A\frac{G^2}{G^2 + q_K^2}
\label{eqn:Kerker}
\end{equation}
mitigates this problem by suppressing the contribution of the
problematic small $G$ components in determining the next input density.
The metric
\begin{equation}
\tilde{M}(\vec{G}) = \frac{G^2 + q_M^2}{G^2}
\label{eqn:Metric}
\end{equation}
enhances the contribution of small $G$ components in
the residual norm, thereby prioritizing the optimization
of these components when determining $n\super{opt}\sub{in}$.
The wavevectors $q_K$ and $q_M$, which control the balance between
short and long-wavelength components, and the prefactor $A$,
which sets the maximum fraction of $n\sub{out}$ that
contributes to the next $n\sub{in}$, can be adjusted
to optimize the convergence of the SCF method.
See Ref.~\citenum{DIIS} for a detailed discussion
of the Pulay-Kerker SCF approach and its performance for
conventional fixed electron number (canonical) DFT calculations.

The $G=0$ component of the electron density, which equals $N/\Omega$,
where $\Omega$ is the unit cell volume, remains fixed in canonical
DFT calculations and is therefore excluded from the metric and the preconditioner.
This is no longer true in fixed-potential DFT, where the number of electrons changes.
With the above prescriptions, the Kerker preconditioner (\ref{eqn:Kerker})
$\rightarrow 0$ as $G\rightarrow 0$ which will prevent the electron number
from changing between SCF iterations.
Likewise, the $G\rightarrow 0$ divergence of the Pulay metric (\ref{eqn:Metric})
cause the residual norm to become undefined when the electron number changes between iterations.

In order to generalize the Pulay-Kerker SCF approach for fixed-potential DFT,
we therefore need to fix the $G\rightarrow 0$ behavior of both the preconditioner and the metric.
The need for the preconditioner and the metric arose from the reciprocal space
Coulomb operator $4\pi/G^2$ in the Hartree potential.
In a uniform electrolyte with Debye screening, the reciprocal space
Coulomb operator instead takes the form $4\pi/(\epsilon_b G^2 + \kappa^2)$
(from (\ref{eqn:LinearPBE}) in reciprocal space with $s=1$).
Since the electrolyte is responsible for fixing the indeterminacy of
the $G=0$ component of the potential, a reasonable \emph{ansatz} for
extending Pulay-Kerker to fixed potential calculations is
replacing $G^2$ with $G^2 + q_\kappa^2$, where
\begin{equation}
q_\kappa = \frac{\kappa}{\sqrt{\epsilon_b}}.
\label{eqn:qKappaOpt}
\end{equation}
In Section~\ref{sec:ResultsSCF} below, we show that setting
\begin{equation}
\tilde{K}(\vec{G}) = A\frac{G^2 + q_\kappa^2}{G^2 + q_\kappa^2 + q_K^2}
\label{eqn:KerkerGC}
\end{equation}
and
\begin{equation}
\tilde{M}(\vec{G}) = \frac{G^2 + q_\kappa^2 + q_M^2}{G^2 + q_\kappa^2}
\label{eqn:MetricGC}
\end{equation}
indeed makes the grand canonical self-consistent field (GC-SCF) method function efficiently,
with optimum convergence for $q_\kappa$ given by (\ref{eqn:qKappaOpt}).

\subsection{Variational minimization: auxiliary Hamiltonian method}\label{sec:AlgoAux}

An alternate approach to solving the Kohn-Sham equations is to
directly minimize the total (free-)energy functional (\ref{eqn:DFT-HKM}),
with $A\sub{HKM}$ given by (\ref{eqn:DFT-KS},\ref{eqn:DFT-ni}),
in terms of the Kohn-Sham orbitals as independent variables.
For joint density-functional theory, this amounts to
\begin{multline}
A = \min_{\{\psi_i(\vec{r}),f_i\}}
\left[
	\sum_i \bigg( \frac{f_i}{2} \int d\vec{r} |\nabla\psi_i(\vec{r})|^2 - TS(f_i) \bigg)
\right. \\ \left.
	+ E_H[n] + E\sub{XC}[n] + A\sub{diel}[n]
	+ \int d\vec{r} V(\vec{r}) n(\vec{r})
\right],
\label{eqn:JDFT-direct}
\end{multline}
where $n(\vec{r})$ is now derived from $\{\psi_i\}$ and
$\{f_i\}$ as given by (\ref{eqn:DensityConstraint}).
In the above minimization, the orbitals $\{\psi_i\}$ must be orthonormal,
the occupation factors must satisfy $0 \le f_i \le 1$, and optionally,
$\sum_i f_i = N$ for the canonical fixed electron number case.

For insulators at $T=0$, the occupations $f_i$ are known in advance.
The constrained optimization over orthonormal $\{\psi_i\}$ is
most efficiently carried out using a preconditioned conjugate-gradients (CG)
algorithm on unconstrained orbitals using the analytically-continued
energy functional approach.\cite{AnalyticContinuedDFT}
Briefly, this approach evaluates the energy functional (\ref{eqn:JDFT-direct})
on a set of orthonormal orbitals, which are a functional of
the unconstrained orbitals used for minimization.
(See Ref.~\citenum{AnalyticContinuedDFT} for further details.)

The general case of metallic systems and/or finite $T$
additionally requires the optimization of $\{f_i\}$, which is
challenging for non-linear optimization algorithms
because of the inequality constraints $0 \le f_i \le 1$.
One possibility is to update the fillings from the Kohn-Sham eigenvalues
using (\ref{eqn:FermiOccupations}) after every few steps of the
CG algorithm,\cite{AlgebraicDFT} but this hinders the convergence of CG
because the functional effectively changes each time the fillings are altered.
The `Ensemble DFT' approach\cite{EnsembleDFT} rectifies this convergence issue
by optimizing the occupation factors at fixed orbitals in an inner loop,
and performing the optimization of orbitals in an outer loop using the CG method,
but this increases the computational cost compared to the case of the insulators.
The SCF approach is typically much more computational efficient than these variants
of the direct variational minimization algorithm with variable occupations.\cite{DIIS}

An alternate strategy for direct variational minimization with
variable occupations is to introduce an auxiliary subspace Hamiltonian
matrix $H\sub{aux}$ as an independent variable,\cite{AuxiliaryHamiltonian}
and setting the occupations $f_i = f(\eta_i)$ in terms of the eigenvalues $\{\eta_i\}$
of $H\sub{aux}$ instead of the Kohn-Sham eigenvalues $\{\varepsilon_i\}$.
(Here, $f(\eta)$ is the Fermi function given by (\ref{eqn:FermiOccupations}),
and the electron chemical potential $\mu$ is chosen so as to satisfy
the electron number constraint $\sum_i f(\eta_i) = N$.)
This eliminates the problematic inequality constraints on the occupations,
and upon minimization, the auxiliary subspace Hamiltonian $H\sub{aux}^{ij}$ approaches the
true subspace Hamiltonian, $H\sub{sub}^{ij} = \langle\psi_i|\hat{H}\sub{KS}|\psi_j\rangle$,
where $\hat{H}\sub{KS} = -\nabla_i^2/2 + V\sub{KS}(\vec{r})$ is the Kohn-Sham Hamiltonian.
By choosing the undetermined unitary rotations of the orbitals $\{\psi_i\}$
to diagonalize $H\sub{aux}$, Ref.~\citenum{AuxiliaryHamiltonian} further shows that
the gradient of the free energy with respect to the independent variables simplifies to
\begin{equation}
\frac{\delta A}{\delta \psi_i(\vec{r})} = f(\eta_i) \bigg(
	\hat{H}\sub{KS}\psi_i(\vec{r}) - \sum_j \psi_j(\vec{r}) H\sub{sub}^{ji}
\bigg)
\end{equation}
and
\begin{multline}
\frac{\partial A}{\partial H\sub{aux}^{ij}} = 
  \delta_{ij} (H\sub{sub}^{ii} - \eta_i) \frac{\partial f(\eta_i)}{\partial\eta_i}\\
- \delta_{ij} \frac{\partial \mu}{\partial\eta_i} \sum_k (H\sub{sub}^{kk} - \eta_k)\frac{\partial f(\eta_k)}{\partial\eta_k}\\
+ (1 - \delta_{ij}) H\sub{sub}^{ij} \frac{f(\eta_i)-f(\eta_j)}{\eta_i-\eta_j}.
\label{eqn:HauxGradient}
\end{multline}
These gradients are used to perform line minimization along a
search direction in the space of independent variables.
With every update of the independent variables,
the orbitals are re-orthonormalized and the unitary rotations of the
orbitals are updated to keep the auxiliary Hamiltonian diagonal.

In the preconditioned CG algorithm, the next search direction
is obtained as a linear combination of the current search direction
and the preconditioned gradients given by
\begin{equation}
K_{\psi_i(\vec{r})} = \hat{T}\sub{inv}\frac{\delta A}{\delta \psi_i(\vec{r})}
\end{equation}
and
\begin{equation}
K_{H\sub{aux}^{ij}} = -K (H\sub{sub}^{ij} - \eta_i\delta_{ij}),
\end{equation}
where $\hat{T}\sub{inv}$ and $K$ are preconditioners.
The role of preconditioning is to balance the weight of different directions
in the minimization space in the explored search directions, and ideally the preconditioner
equals the inverse of the Hessian (which is difficult to compute exactly).
For the orbital directions, the standard preconditioner $\hat{T}\sub{inv}$
resembles the inverse of the dominant kinetic energy operator in
the Kohn-Sham Hamiltonian.\cite{AnalyticContinuedDFT}
For the auxiliary Hamiltonian direction, the preconditioner removes the Fermi
function derivatives and finite difference factors from (\ref{eqn:HauxGradient})
in order to more equitably weight all components of $H\sub{aux}$.
The preconditioning factor $K$ controls the overall contribution of the $H\sub{aux}$ components
relative to that of the orbital components, and is adjusted to achieve optimum convergence.

In this auxiliary Hamiltonian (AuxH) approach, the orbitals and occupations
are continuously and simultaneously optimized to minimize the total free energy,
resulting in better convergence and computational efficiency comparable
to the fixed-occupations insulator case, and competitive with
the SCF method even for metallic systems.
See Ref.~\citenum{AuxiliaryHamiltonian} for further details on the algorithms
and performance comparisons for conventional fixed electron number calculations.

Now, for fixed potential calculations, we set the occupations to Fermi functions
of the auxiliary Hamiltonian eigenvalues at a specified $\mu$,
instead of selecting $\mu$ based on the electron number constraint.
Correspondingly, the second term of the auxiliary Hamiltonian gradient
(\ref{eqn:HauxGradient}), which arises from this constraint,
drops out, and the algorithm requires no further modification.

The convergence rate of this algorithm, however, is sensitive to the preconditioning factor
$K$ and we propose a modified heuristic to update $K$ automatically and continuously.
At the end of each line minimization, the derivative of the
optimized free energy $A\sub{min}$ with respect to $K$ can be
evaluated from the overlap between the auxiliary Hamiltonian
gradient and search direction.\cite{AuxiliaryHamiltonian}
The line-minimized energy is optimum for $\partial A\sub{min}/\partial K = 0$.
Therefore, if variation of $A\sub{min}$ with respect to $K$ is convex,
we should increase $K$ if we find $\partial A\sub{min}/\partial K < 0$ and vice versa.
However convexity is often lost if $K$ is initialized at too high a value.
Therefore, our heuristic tries to zero $\partial A\sub{min}/\partial K$
while limiting the contribution of the auxiliary Hamiltonian gradient.
In particular, we update
\begin{equation}
K \leftarrow K \times \max\left[
\exp\left(f\sub{sat}\left( \frac{-\partial A\sub{min}/\partial K}{g\sub{tot}} \right)\right),
\frac{g\sub{tot}}{2g\sub{aux}} \right]
\label{eqn:Kadjust}
\end{equation}
at the end of each line minimization,
where $f\sub{sat}(x)\equiv x/\sqrt{1+x^2}$ to saturate the factor
by which $K$ can change in one iteration,
$g\sub{aux}$ is the overlap of the $H\sub{aux}$ components of the
gradient and preconditioned gradient, and $g\sub{tot}$ is the total
overlap of the gradient and preconditioned gradients (orbital + $H\sub{aux}$).
Finally, we reset the conjugate-gradient algorithm (i.e. set the search
direction to the negative of the preconditioned gradient direction)
after $K$ has increased or decreased by a factor greater than $e^2$.
We do this because dynamically changing the preconditioner technically invalidates
the strict orthogonality of the CG search direction with previous directions.
Section~\ref{sec:ResultsAux} below shows that this heuristic exceeds
the convergence obtained with fixed $K$, while section~\ref{sec:AlgoCompare}
shows that the grand-canonical auxiliary Hamiltonian (GC-AuxH) algorithm
consistently outperforms GC-SCF for fixed-potential calculations.

\section{Results}
\makeatletter{}\subsection{Computational details}\label{sec:CompDetails}

We implement all algorithms and perform all calculations using the
open-source plane-wave density-functional theory software, JDFTx.\cite{JDFTx}
Below, we specify computational and convergence parameters in
atomic units (distances in bohrs $a_0 \approx 0.529$~\AA~ and
energies in Hartrees $E_h\approx 27.2$~eV), but present any
physically relevant properties in conventional units (\AA, eV).
All calculations in this work employ the PBE\cite{PBE} exchange-correlation functional
with GBRV ultrasoft pseudopotentials\cite{GBRV} at a kinetic energy cutoff of
$20~E_h$ for Kohn-Sham orbitals and $100~E_h$ for the charge density.
The metal surface calculations use inversion-symmetric slabs of at least five layers,
with at least 15~\Angstrom~vacuum separation and truncated Coulomb
potentials\cite{TruncatedEXX} to minimize interactions with periodic images.
For Brillouin zone integration, we use a Fermi smearing of $0.01~E_h$
and a Monkhorst-Pack $k$-point mesh along the periodic directions with
the number of $k$-points chosen such that the effective supercell
is larger than 30~\AA~ in each direction.
We use the CANDLE solvation model to describe the effect of liquid water
and Debye screening due to 1M electrolyte, which we showed recently to most
accurately capture the solvation of highly-charged negative and positive solutes.\cite{CANDLE}
We emphasize that the methods and algorithms described above do not rely on specific choices for
the pseudopotential, exchange-correlation functional, $k$-mesh or solvation model;
we keep these computational parameters constant here for consistency.
 
\makeatletter{}\subsection{Convergence of the GC-SCF method}\label{sec:ResultsSCF}

\begin{figure}
\includegraphics[width=\columnwidth]{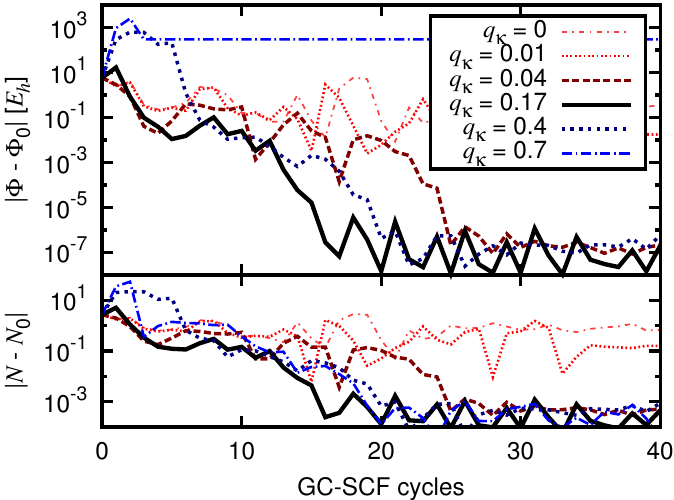}
\caption{
Dependence of GC-SCF convergence on low-frequency cutoff wavevector $q_\kappa$.
The upper panel shows the convergence of the grand free energy $\Phi$ (towards its final value $\Phi_0$) on
a logarithmic scale, and the lower panel shows that of the electron number $N$ (towards its final value $N_0$).
Best convergence is obtained with $q_\kappa = 0.17 a_0^{-1}$,
the inverse Debye-screening length, which we use as the default value henceforth.
Results shown here are for a five-layer (5ML) Cu(111) slab solvated in
1M CANDLE aqueous electrolyte, with potential fixed to 1V SHE ($\mu=-0.208 E_h$),
starting from a converged neutral calculation of the same slab in vacuum.
\label{fig:SCF-qKappa}}
\end{figure}

The self-consistent field (GC-SCF) algorithm summarized in section~\ref{sec:AlgoSCF}
depends on several parameters that control its iterative convergence.
All these parameters are common to the conventional fixed-charge version (SCF)
and the fixed-potential variant (GC-SCF) introduced here, except for the
low-frequency cutoff wavevector $q_\kappa$ that is necessary to
allow the net electron number to change in the fixed-potential case.
Figure~\ref{fig:SCF-qKappa} compares the dependence of GC-SCF convergence on
$q_\kappa$ for a prototypical calculation of an electrochemical system:
a Cu(111) surface treated using a five-layer inversion-symmetric slab
surrounded by 1M aqueous non-adsorbing electrolyte treated using
the CANDLE solvation model,\cite{CANDLE} at a fixed potential of $\mu=-0.208 E_h$ (1V~SHE).
(The remaining GC-SCF parameters are set to their default values which we discuss below.)
The best convergence is obtained with $q_\kappa = 0.17$ which corresponds
to the Debye screening length of the electrolyte (\ref{eqn:qKappaOpt}).
For small $q_\kappa$, including the conventional case of $q_\kappa = 0$,
the number of electrons does not respond sufficiently quickly, stalling at about
0.1~electrons from the converged value, correspondingly with the free energy 
stalling at about 0.1~$E_h$ ($\approx 2.7$~eV) away from the converged value.
Larger $q_\kappa$ causes the electron number to change too rapidly,
hindering convergence and eventually leading to a divergence
as seen for the case of $q_\kappa = 0.7$.
An issue remains in the convergence independent of $q_\kappa$:
after initial convergence, the free energy oscillates at the $10^{-6}~E_h$
level, while the electron number oscillates at the $10^{-3}$ level.

\begin{figure}
\includegraphics[width=\columnwidth]{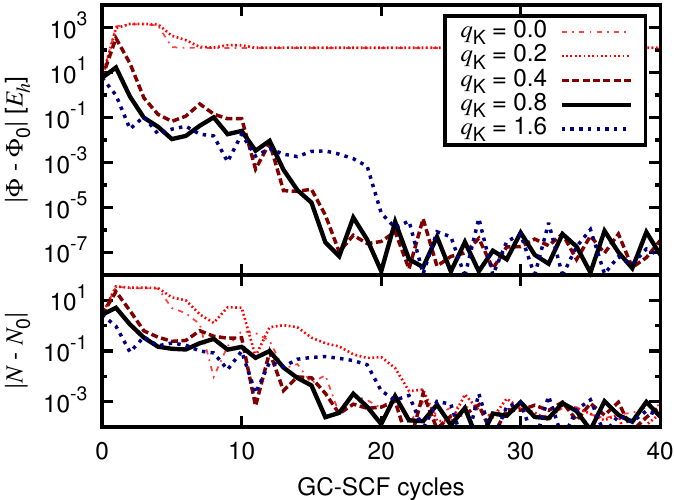}
\caption{
Dependence of GC-SCF convergence on Kerker-mixing wavevector $q_K$.
Good convergence is observed near the typical recommended value
$q_K = 0.8~a_0^{-1}$ ($\approx 1.5$~\AA\super{-1}).
Convergence is relatively insensitive to $q_K$ near this value, but becomes
unstable for small $q_K$ approaching the low-frequency cutoff $q_\kappa$.
System and remaining details are identical to Figure~\ref{fig:SCF-qKappa}.
\label{fig:SCF-qKerker}}
\end{figure}

Keeping $q_\kappa$ at this optimum value given by (\ref{eqn:qKappaOpt}),
we next examine the dependence of GC-SCF convergence on the remaining
algorithm parameters for the same example system.
Figure~\ref{fig:SCF-qKerker} shows the dependence on the Kerker mixing wavevector $q_K$,
which helps stabilize the GC-SCF algorithm against long-wavelength charge oscillations.
Optimal convergence is obtained for the typical recommended value\cite{DIIS}
of $0.8~a_0^{-1}$ ($\approx 1.5$~\AA\super{-1}).
As expected, convergence is relatively insensitive to the exact choice of $q_K$,
as long as $q_K$ does not become so small that convergence is ruined by charge sloshing.
Notice that the final convergence beyond the $10^{-6}~E_h$ and $10^{-3}$ electron
level remains an issue that is not resolved for any choice of $q_K$.

\begin{figure}
\includegraphics[width=\columnwidth]{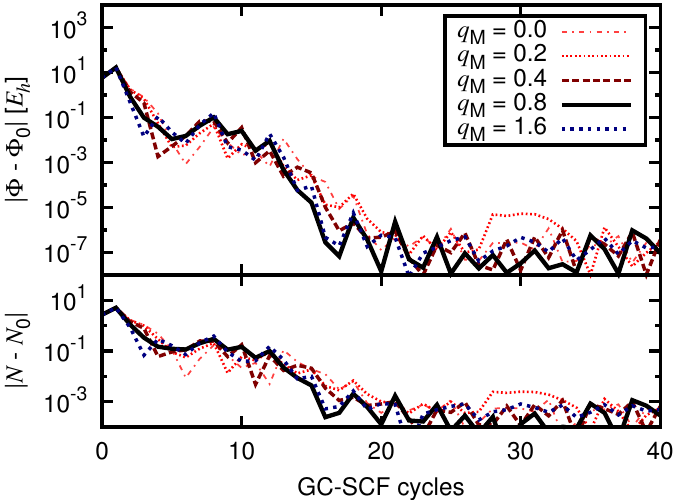}
\caption{
Dependence of GC-SCF convergence on Pulay-metric wavevector $q_M$.
Convergence is relatively insensitive to $q_M$,
and we set $q_M = q_K = 0.8~a_0^{-1}$ henceforth.
System and remaining details are identical to Figure~\ref{fig:SCF-qKappa}.
\label{fig:SCF-qMetric}}
\end{figure}

Next, Figure~\ref{fig:SCF-qMetric} shows the variation of GC-SCF convergence
with the wavevector $q_M$ controlling the reciprocal-space metric used by the Pulay algorithm.
The convergence is entirely insensitive to this choice, and we henceforth set
$q_M = q_K = 0.8~a_0^{-1}$ (the recommended value\cite{DIIS}).
Again, the final convergence issue remains unaffected by the choice of $q_M$.

\begin{figure}
\includegraphics[width=\columnwidth]{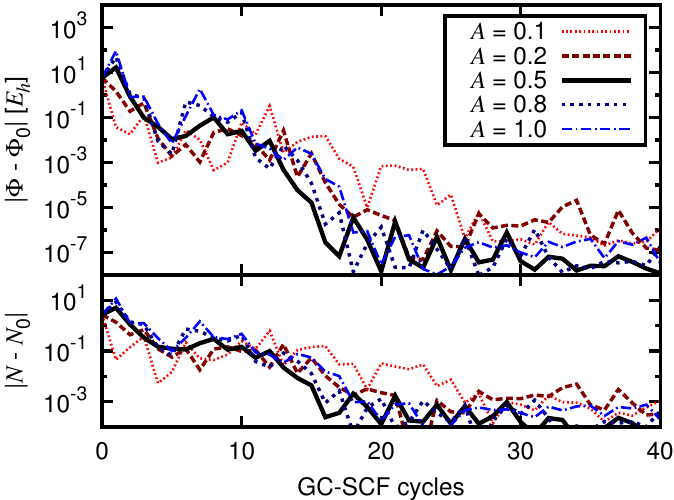}
\caption{
Dependence of GC-SCF convergence on maximum Kerker-mixing fraction $A$.
Convergence is relatively insensitive to $A$, except that it slows down for small $A$
We henceforth set $A = 0.5$ which nominally exhibits the best convergence.
System and remaining details are identical to Figure~\ref{fig:SCF-qKappa}.
\label{fig:SCF-mixFraction}}
\end{figure}

Finally, Figure~\ref{fig:SCF-mixFraction} compares the dependence of GC-SCF convergence
on the maximum Kerker mixing fraction $A$, which effectively controls what fraction
of the new electron density is mixed into the current value.
We find nominally best convergence for $A = 0.5$, but the performance of
other values is not much worse.
Smaller values of $A$ lead to greater stability initially, but marginally slower convergence later on,
while larger values of $A$ lead to greater oscillations initially, but faster convergence later on.
Regardless, as before, good convergence is obtained until the free energy reaches
the $10^{-6}~E_h$ level, but continues to oscillate at that level beyond that point.

\begin{figure}
\includegraphics[width=\columnwidth]{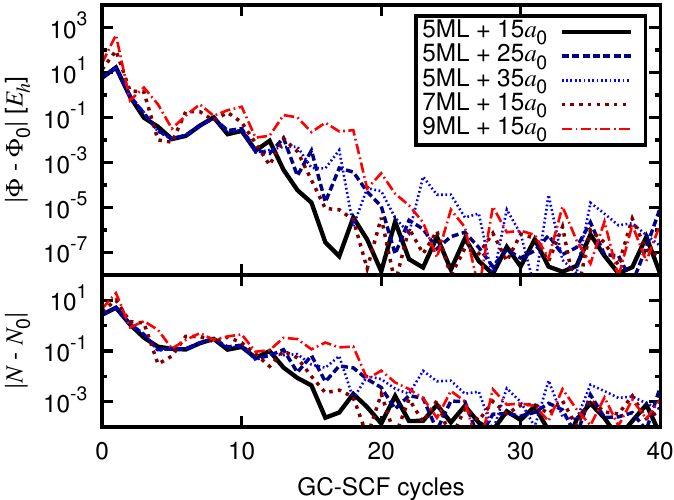}
\caption{
Variation of GC-SCF convergence using the default parameters determined above
with system size: both with number of slab layers ranging from 5ML to 9ML
with fixed vacuum spacing 15~$a_0$, and with vacuum spacing ranging from
15~$a_0$ to 35~$a_0$ with the 5ML slab.
Convergence slows only marginally with increasing system size,
with either vacuum spacing or layer count.
Each calculations is for solvated Cu(111) charged to 1V~SHE in CANDLE electrolyte,
starting from the state of the corresponding converged neutral vacuum calculation.
\label{fig:SCF-size}}
\end{figure}

This final convergence issue may not affect practical calculations where
relevant energy differences are at the $10^-3~E_h$ level or higher.
However, smooth exponential convergence to the final answer is desirable
as this makes it easier to determine when the target accuracy has been reached.
Unfortunately, no combination of GC-SCF parameters achieves
uniformly smooth convergence for the fixed-potential case.
On the other hand, with our $q_\kappa$ modification, the GC-SCF algorithm
at least converges to the $10^{-6}~E_h$ level independent of system size
(Figure~\ref{fig:SCF-size}), with the number of cycles required for convergence
remaining mostly unchanged with increasing number of Cu(111) layers,
and with increasing thickness of the solvent region.

Note that the standard fixed-charge SCF method converges the energy
smoothly and exponentially in most cases, including in our implementation in JDFTx.
Our implementation of the GC-SCF method in JDFTx uses exactly the same code, except
for the preconditioner and metric modifications (due to $q_\kappa$) presented here.
Therefore we believe that the final convergence difficulty in GC-SCF is
a property of the algorithm itself, rather than an implementation issue.

\subsection{Convergence of the GC-AuxH method}\label{sec:ResultsAux}

\begin{figure}
\includegraphics[width=\columnwidth]{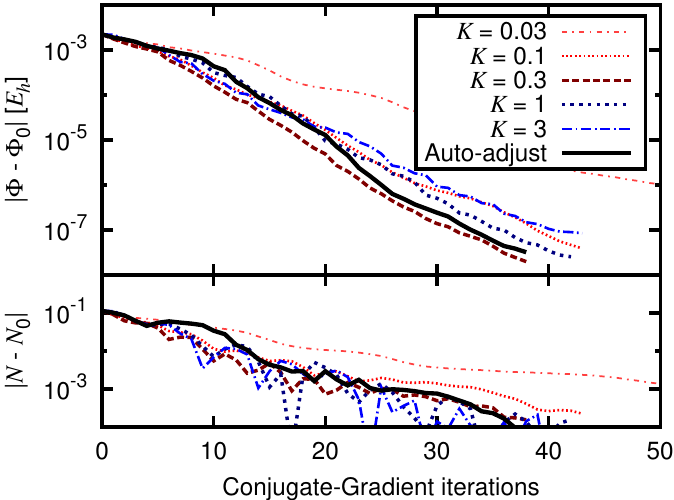}
\caption{
Dependence of the convergence of the GC-AuxH variational-minimize method
on $K$, the preconditioning scale factor for subspace rotations
generated by the auxiliary Hamiltonian.
Near-optimal convergence is obtained when $K$ is automatically adjusted using
the heuristic given by (\ref{eqn:Kadjust}), which we use by default henceforth.
Calculations are for solvated 5ML Cu(111) at 1V~SHE,
starting from the corresponding neutral vacuum calculation,
exactly as in Figures~\ref{fig:SCF-qKappa}-\ref{fig:SCF-mixFraction}.
Note the smooth exponential convergence (without oscillations in
electron number and free energy) here, in contrast to the GC-SCF case.
\label{fig:Aux-K}}
\end{figure}

The grand-canonical auxiliary Hamiltonian (GC-AuxH) approach discussed in section~\ref{sec:AlgoAux},
directly minimizes the total free energy of the system without assuming any
models for physical properties of the system (such as dielectric
response models that are built into the SCF mixing schemes).
This algorithm contains a single parameter $K$, which weights the
relative contributions of the Kohn-Sham orbital and subspace Hamiltonian degrees
of freedom in the conjugate-gradients search direction for free energy minimization.

Figure~\ref{fig:Aux-K} shows the dependence of iterative convergence of the GC-AuxH algorithm
on this preconditioning parameter $K$ for the same Cu(111) test problem considered above.
If the preconditioning factor $K$ is held fixed, the rate of convergence is sensitive
to the choice of $K$, with the optimum choice being $K \approx 0.3$ for this system.
With the preconditioner auto-adjusted using the heuristic given by (\ref{eqn:Kadjust}),
we find that indeed the convergence picks up from that of
the sub-optimal $K=1$ towards that of the optimal value.
More importantly, we observe smooth exponential convergence
of both the free energy and the electron number,
in contrast to our experiences with the GC-SCF method.

\begin{figure}
\includegraphics[width=\columnwidth]{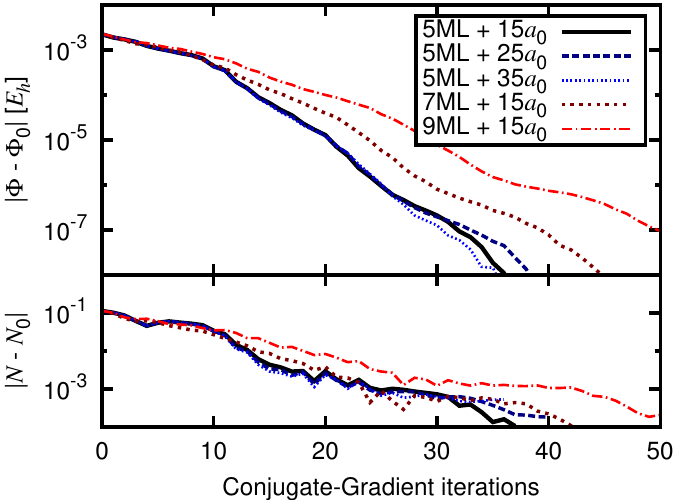}
\caption{
Variation of convergence of the GC-AuxH variational-minimize method with Cu(111)
system size (exactly analogous to Figure~\ref{fig:SCF-size} for the GC-SCF method).
Convergence is invariant with vacuum spacing, and slows down
only marginally with number of layers.
\label{fig:Aux-size}}
\end{figure}

Figure~\ref{fig:Aux-size} further shows that this smooth convergence sustains with changing system size.
In particular, the convergence is virtually unchanged with the thickness of the solvent regions,
but slows down slightly with increasing number of copper layers in the surface slabs.

\subsection{Comparison of algorithms}\label{sec:AlgoCompare}

Having analyzed and optimized the convergence of the GC-SCF and GC-AuxH methods,
we now compare the performance of these algorithms for a few different cases.
In this comparison, we also include the present state of the art: the `Loop' method
which uses a secant method to adjust the number of electrons in an outer loop
to match the specified electron chemical potential.\cite{UadjustLoop}
For a fair comparison, we use the fixed-charge SCF method in the inner loops,
because it achieves the fastest convergence; this algorithm works equally
well with the AuxH method in the inner loop, but is then marginally slower.
In each test case, we start all three algorithms from the same starting point:
the converged state of the corresponding neutral (fixed-charge) calculation.
Different test cases effectively perturb the potential by different amounts, thereby
testing the algorithms for a range of proximities between initial and final states.
Also, we now compare the wall time between algorithms, because there is no
straightforward correspondence between GC-SCF cycles and conjugate-gradient
iterations of the GC-AuxH method.
The relative wall-time performance of these fairly distinct algorithms
will depend to an extent on details of code optimization for each.
However, there is no perfect metric for comparing these algorithms
and wall time suffices for a rough qualitative comparison.

\begin{figure}
\includegraphics[width=\columnwidth]{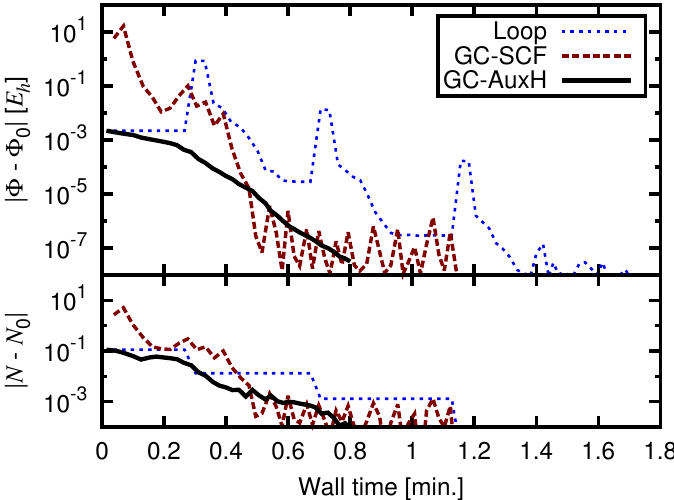}
\caption{
Performance comparison of the new GC-SCF and GC-AuxH methods
with the previous state of the art: a `Loop' over fixed-charge calculations
(using the secant method to adjust the charge to match the potential).
Both new methods reach $10^{-7}~E_h$ free energy accuracy in half the wall time of the
Loop method, but the GC-AuxH method is the clear winner with smooth exponential convergence.
Calculations here are for the 5ML Cu(111) slab at 1V~SHE, as before.
Timings are measured on a single 32-core NERSC Cori node in all cases.
\label{fig:Compare-Cu5}}
\end{figure}

First, Figure~\ref{fig:Compare-Cu5} compares the performance of the algorithms
for the 5-layer Cu(111) slab used in all the tests so far.
The spikes in the free energy seen in the Loop method
are the points where the electron number changes in the outer loop
and a new SCF convergence at fixed charge begins.
Both the GC-SCF and GC-AuxH methods are quite competitive,
cutting the time to convergence within $10^{-6}~E_h$ in half
compared to the Loop method.
Given the smooth convergence beyond $10^{-6}~E_h$ however,
the GC-AuxH method is preferable over GC-SCF for fixed potential calculations.
Note that in the fixed-charge case, when SCF converges smoothly,
it often outperforms the AuxH method as mentioned above.
The advantage of the AuxH and GC-AuxH method is their stability on account of being
variational methods: the free energy is guaranteed to decrease at every step.
Consequently, the convergence difficulties of the GC-SCF method make
the variationally stable GC-AuxH method relatively more attractive.

\begin{figure}
\includegraphics[width=\columnwidth]{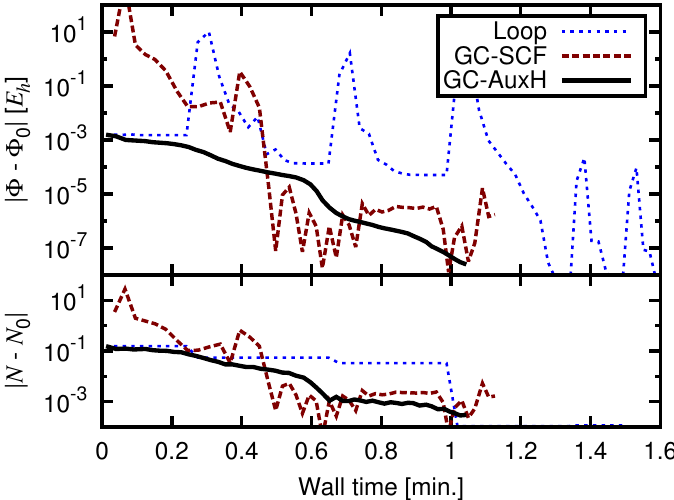}
\caption{
Similar to Figure~\ref{fig:Compare-Cu5}, but for a 5ML Pt(111) slab at 0V~SHE ($\mu=-0.171~E_h$).
Platinum is neutral at $\approx 0.8$V~SHE, so this corresponds to negatively
charging the slab, which activates CANDLE's asymmetry correction.
Additionally, platinum has $d$ bands crossing the Fermi level
in contrast to copper which has occupied $d$ bands.
This calculation therefore explores a more complicated charge vs.
potential landscape, causing deviations from exponential convergence,
but the relative performance of the algorithms remains similar.
\label{fig:Compare-Pt5}}
\end{figure}

Next, we compare these algorithms for more complex text cases.
Figure~\ref{fig:Compare-Pt5} compares the convergence for a
five-layer Pt(111) slab fixed to a potential of $\mu=-0.171~E_h$ (0V~ SHE).
At this potential, the surface of Pt(111) charges negatively,
and the CANDLE solvation model brings the cavity closer to the electrons
to capture the more effective solvation of negative charges by liquid water.
Additionally, the dielectric response of platinum is more complex than copper
due to the partially filled $d$ shell.
Both of these factors make this system harder to converge than the previous test case,
and therefore the convergence of the GC-AuxH method is no longer clearly a single exponential.
Despite this, both the direct grand-canonical methods converge faster than the Loop method,
with the GC-AuxH method eking out an advantage in final convergence as before.

\begin{figure}
\includegraphics[width=\columnwidth]{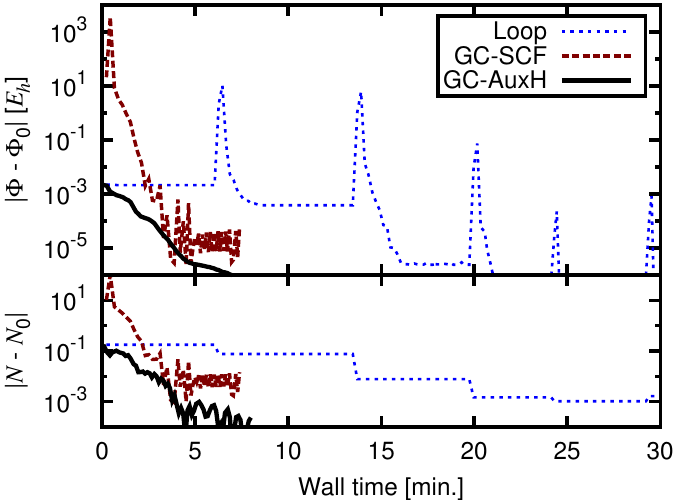}
\caption{
Similar to Figures~\ref{fig:Compare-Cu5} and \ref{fig:Compare-Pt5},
but for a 5ML Pt(111) slab decorated with a $\sqrt{3}\times\sqrt{3}$
partial monolayer of adsorbed chloride anions (1/3 coverage) at 0V~SHE.
The increased complexity slows down the convergence of the GC-SCF
method, which also slows down the Loop over fixed-charge SCF calculations,
while the GC-AuxH method continues to exhibit rapid near-exponential convergence
with no charge oscillations, beating the Loop method by a factor of 4 in time.
All subsequent calculations of complex metal surfaces with adsorbates
therefore use the GC-AuxH method.
\label{fig:Compare-Pt5_Cl}}
\end{figure}

Finally, Figure~\ref{fig:Compare-Pt5_Cl} compares the convergence
for chloride anions adsorbed at one-third monolayer coverage,
in a $\sqrt{3}\times\sqrt{3}$ supercell of a five-layer Pt(111) slab.
Despite the increased complexity, the direct grand-canonical methods
exhibit the best convergence, edging out the Loop method by a factor of
four in wall time now, again with smoothest convergence for the GC-AuxH method.
Due to the systematic convergence advantage of the GC-AuxH method, we use it
for all remaining calculations in this work and recommend it as the default
general purpose algorithm for converging electrochemical calculations.

In the theory section and all calculations so far,
we used Fermi smearing where the electron occupations
are given by the Fermi function (\ref{eqn:FermiOccupations}).
Practical $k$-point meshes typically require the use of a temperature $T$
substantially higher than room temperature (we used 0.01~$E_h$ which is
approximately ten times higher), which could result in inaccurate free energies.
Such errors can be reduced substantially by changing the functional form of
the occupations and the electronic entropy, with the caveat that the
smearing width $T$ no longer corresponds to an electron temperature.
Common modifications include Gaussian smearing, where the Fermi functions
are replaced with error functions, and Cold smearing,\cite{ColdSmearing} where the
functional form is chosen to cancel the lowest order variation of the free energy with $T$.
(See Ref.~\citenum{ColdSmearing} for details.)

\begin{figure}
\includegraphics[width=\columnwidth]{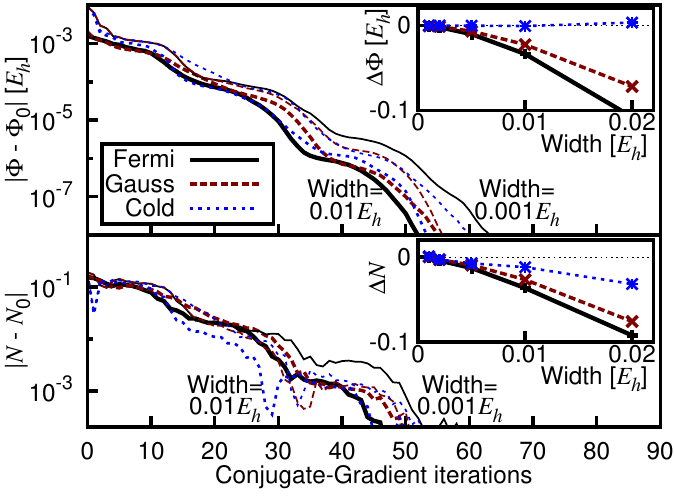}
\caption{
Comparison of the convergence of the GC-AuxH method for the
5ML Pt(111) slab at 0V~SHE using different smearing functions.
Insets show the variation of converged properties with smearing width.
Cold smearing\cite{ColdSmearing} substantially reduces the finite-width error in the free energy,
and to a lesser extent in the electron number, compared to Gaussian and Fermi smearing.
At the same width of 0.01~$E_h$, the GC-AuxH method with cold smearing converges
marginally slower than Gaussian or Fermi smearing, but still faster than
when using any of the smearing methods with the smaller width of 0.001~$E_h$.
\label{fig:Smearing}}
\end{figure}

Figure~\ref{fig:Smearing} compares the performance of the preferred GC-AuxH method for various smearing methods.
The insets show the variation of the converged free energy and electron number with smearing width $T$.
Gaussian smearing reduces the coefficient of the quadratic $T$ dependence compared to Fermi smearing,
while Cold smearing cancels the quadratic dependence altogether, by design.
The variation of electron number with smearing width is also reduced by Cold smearing, but to a lesser extent.
Notice that the use of Cold smearing marginally slows down the iterative convergence of the GC-AuxH method.
However, using Cold smearing at a high width of 0.01~$E_h$ is still faster than using
any smearing method with the lower width 0.001~$E_h$ that is close to room temperature
(because of the far denser Brillouin zone sampling required for smaller widths).
Therefore, it is still advantageous to use Cold smearing at elevated smearing widths,
and so we use Cold smearing with a width of 0.01~$E_h$ for the final demonstration below.
 
\makeatletter{}\subsection{Under-potential deposition of Cu on Pt(111)}\label{sec:UPD}

The application of sufficiently negative (reductive) potentials on an electrode
immersed in a solution containing metal ions, reduces those ions and results in
bulk electro-deposition of metal on the surface. Additionally, for many pairs of metals,
a single monolayer of one metal deposits on a surface of the other at an
under potential, that is, at a potential less favorable than for bulk deposition.
This phenomenon of under-potential deposition (UPD) has several technological applications
since it enables precise synthesis of heterogeneous metal interfaces.
It also serves as an archetype for fundamental studies of electrochemical processes
(see Ref.~\citenum{UPD-Review} for an extensive review), which makes it a perfect example
for demonstrating our grand-canonical density-functional theory method.

The basic reason for underpotential deposition is that the heterogeneous binding
between the two metals is stronger than the homogeneous binding of the depositing
metal to itself. Indeed, metal pairs that exhibit underpotential deposition
also display analogous phenomena in vapor adsorption.\cite{UPD-CuPtVapor}.
However, the process in solution is far more complicated and highly sensitive to the
composition of the solution because of competing adsorbates,\cite{UPD-CompetingAdsorbates}
as well as to the structure of the electrode surface.\cite{UPD-ElectrodeStructureEffects}

\begin{figure}
\includegraphics[width=\columnwidth]{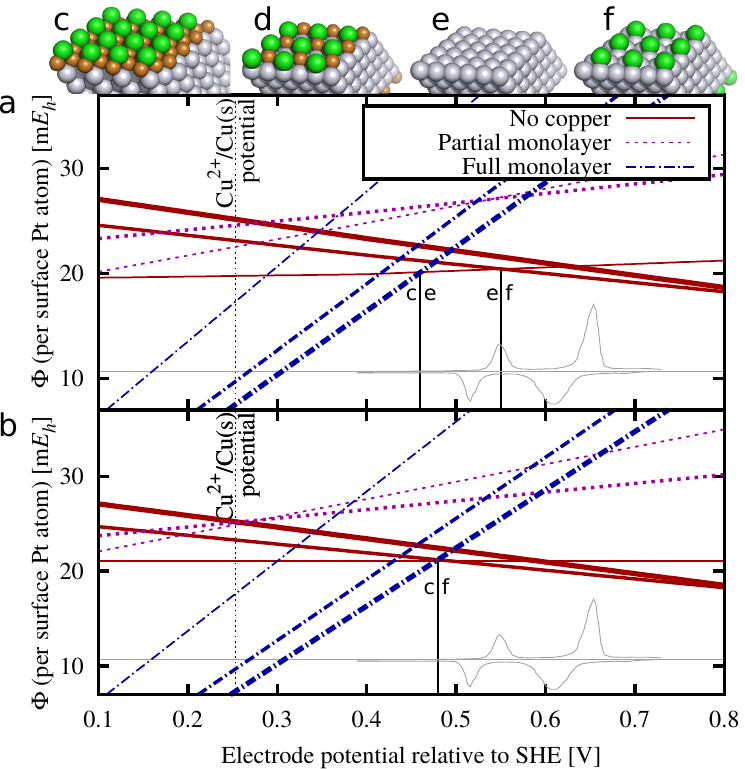}
\caption{
Underpotential deposition of Cu on Pt(111) from an aqueous solution containing
10$^{-3}$~mol/liter Cu\super{2+} ions and 10$^{-4}$~mol/liter Cl$^-$ ions.
Calculated free energies for various adsorbate configurations as a function of electrode potential
are shown as calculated by explicit fixed-potential solvated calculations in (a), and from vacuum
calculations in (b), with the experimental voltammogram\cite{CuPtCl-FullMonolayer-LEED1} shown for comparison.
Solid red lines indicate no copper, purple dotted lines indicate $2\times 2$ partial copper monolayer
and blue dot-dash lines indicate full copper monolayer.
Thickness of the lines indicate Cl coverage, ranging from no Cl (thinnest),
$2\times 2$ (1/4) coverage (intermediate) to $\sqrt{3}\times\sqrt{3}$ (1/3) coverage (thickest).
Prominent configurations are sketched in (c-f).
Vacuum calculations in (b) predict only a single voltammetric peak in disagreement with experiment.
Explicit potential-dependent solvated calculations in (a) predict two peaks
in qualitative agreement with experiment ($\sim 0.1$~V accuracy),
with the second peak due to chloride desorption.\cite{
CuPtCl-FullMonolayer-Xrays,CuPtCl-FullMonolayer-LEED1, CuPtCl-FullMonolayer-LEED2}
The partial Cu monolayer (d) proposed by some\cite{CuPtCl-PartialMonolayer-LEEDandXray}
is not predicted to be the most stable configuration at any relevant potential.
\label{fig:UPD}}
\end{figure}

The UPD of copper on Pt(111) in the presence of chloride anions is particularly
interesting and the subject of considerable debate in the literature.
Voltammetry for this system\cite{CuPtCl-FullMonolayer-LEED1} exhibits two well-separated
under-potential peaks, as shown in the background of Figure~\ref{fig:UPD}.
Certain LEED and in situ X-ray scattering studies of this system\cite{CuPtCl-PartialMonolayer-LEEDandXray}
find evidence of a $2\times 2$ bilayer of copper and chloride ions co-adsorbed on the surface
at potentials between the two peaks, suggesting that one peak corresponds
to a formation of a partial layer, and the second peak, to the formation of the full monolayer.
In contrast, other studies\cite{CuPtCl-FullMonolayer-Xrays,CuPtCl-FullMonolayer-LEED1,
CuPtCl-FullMonolayer-LEED2} do not find this signature and propose that the
additional peak arises from adsorption and desorption of chloride ions alone.

To address this debate, we perform grand-canonical density functional theory calculations
of various configurations of copper and chlorine adsorbed on a 5-layer Pt(111) slab
in $\sqrt{3}\times\sqrt{3}$ and $2\times 2$ supercells.
We determine the most stable configurations at each potential($\mu$) and the potentials
at which transitions between configurations occur by comparing their grand free energies.
The relevant grand free energy of a configuration $\alpha$ containing
$N\sub{Pt}^\alpha$ platinum atoms in the slab with $N\sub{Cu}^\alpha$ copper and
$N\sub{Cl}^\alpha$ chlorine atoms adsorbed at the surface within the calculation cell is
\begin{equation}
\tilde{\Phi}^\alpha(\mu) = \frac{\Phi^\alpha(\mu) - \mu\sub{Pt}N\sub{Pt}^\alpha
- \mu\sub{Cu}N\sub{Cu}^\alpha - \mu\sub{Cl}N\sub{Cl}^\alpha }{ N\sub{Pt}\super{surf} },
\label{eqn:FreeEnergyUPD}
\end{equation}
normalized by the number of surface platinum atoms $N\sub{Pt}\super{surf}$ in order
to correctly compare energies of calculations in different supercells.
($N\sub{Pt}\super{surf} = 2$ for the unit cell, 6 for the $\sqrt{3}\times\sqrt{3}$ supercell and
8 for the $2\times 2$ supercell, accounting for the top and bottom surfaces in the inversion-symmetric setup.)
Since no light atoms are present, we safely neglect changes in vibrational
contributions to the free energy between adsorbate configurations.

Above, $\Phi^\alpha(\mu)$ is the free energy of adsorbate configuration $\alpha$ calculated by fixed-potential DFT,
which is grand canonical with respect to the electrons at chemical potential $\mu$ (related to the electrode
potential by (\ref{eqn:muCalibration}) as discussed at the end of Section~\ref{sec:TheoryJDFT}).
Then, (\ref{eqn:FreeEnergyUPD}) above calculates the free energy $\tilde{\Phi}^\alpha(\mu)$
which is additionally grand canonical with respect to all relevant atoms with chemical potentials
$\mu\sub{Pt}$, $\mu\sub{Cu}$ and $\mu\sub{Cl}$.
Several conventions are possible in defining the electron-grand-canonical free energy $\Phi$,
depending on what electron number we subtract: change from neutral value, total electron number,
or number of valence electrons in pseudopotential DFT calculations.
The atom chemical potentials would then respectively correspond to
neutral atoms, bare nuclei or pseudo-nuclei (nuclei + core electrons in pseudopotential).
The full grand canonical free energy $\tilde{\Phi}$ does not depend on this choice.
In our JDFTx implementation, we choose the last option above (number of valence electrons
and correspondingly atom chemical potentials of the pseudo-nuclei).

The bulk of the platinum electrode sets the Pt chemical potential,
$\mu\sub{Pt} = E\sub{Pt(s)} - \mu N^e\sub{Pt(s)}$,
where $E\sub{Pt(s)}$ is the DFT energy of a bulk fcc Pt calculation
with a single atom in the unit cell, and $N^e\sub{Pt(s)}$
is the number of valence electrons in that calculation.
The second term here implements the electron counting convention discussed above.
Next, copper ions in solution set $\mu\sub{Cu}$, but directly calculating
the free energy of such ions using solvation models is error-prone.\cite{CANDLE}
So, instead, we use the DFT calculated energy, $E\sub{Cu(s)}$, of a bulk fcc Cu calculation
(containing $N^e\sub{Cu(s)}$ valence electrons), and relate it to the free energy
of the ion via the experimentally-determined standard reduction potential
$U\sub{Cu\super{2+}$\rightarrow$Cu(s)} = 0.342$~V SHE.\cite{CRC-Handbook}
This yields
\begin{multline}
\mu\sub{Cu} = \left( E\sub{Cu(s)} - \mu N^e\sub{Cu(s)} \right) \\
	+ 2\left( \mu - \mu\sub{SHE} + eU\sub{Cu\super{2+}$\rightarrow$Cu(s)} \right) \\
	+ k_B T \ln [\mathrm{Cu}^{2+}],
\end{multline}
where the second term accounts for the change from Cu(s) to $\mathrm{Cu}^{2+}$ ions,
and the final term accounts for change in ionic concentration from the
standard value of 1~mol/liter to the current value of [Cu$^{2+}$] (in mol/liter).
Similarly, chlorine ions in solution set $\mu\sub{Cl}$, but to minimize DFT errors,
we connect to the DFT calculated energy, $E\sub{Cl(at)}$, of an isolated Chlorine atom
(containing $N^e\sub{Cl(at)}$ valence electrons), via the experimentally-determined
atomization energy $E\sub{Cl\sub{2}$\rightarrow$2Cl(at)} = 242.6$~kJ/mol,\cite{cccbdb}
gas-phase entropy $S\sub{Cl\sub{2}(g)} = 223.1$~J/mol-K,\cite{CRC-Handbook}
and reduction potential $U\sub{Cl\sub{2}(g)$\rightarrow$2Cl$^-$} = 1.358$~V SHE.\cite{CRC-Handbook}
Specifically,
\begin{multline}
\mu\sub{Cl} = \left( E\sub{Cl(at)} - \mu N^e\sub{Cl(at)} \right)
	- \frac{1}{2}\left( E\sub{Cl\sub{2}$\rightarrow$2Cl(at)} + TS\sub{Cl\sub{2}(g)} \right) \\
	- \left( \mu - \mu\sub{SHE} + eU\sub{Cl\sub{2}(g)$\rightarrow$2Cl$^-$} \right) \\
	+ k_B T \ln [\mathrm{Cl}^-],
\end{multline}
where the second term accounts for the change from atomic to gas-phase chlorine,
the third term for the change to chloride ions, and the final term for the change
in chloride ion concentration to [Cl$^-$] (in mol/liter).

Figure~\ref{fig:UPD}(a) shows the calculated grand free energies as a function of electrode
potential for a number of Cu and Cl adsorbate configurations on the surface of Pt(111).
At high potentials, the most stable (lowest free energy) configuration is
1/4 Cl coverage (Figure~\ref{fig:UPD}(f)), which transitions to a clean
Pt surface (Figure~\ref{fig:UPD}(e)) at a potential of 0.55~V SHE.
Upon further lowering the potential, the stable configuration transitions to a full monolayer
of copper with 1/3 Cl coverage (Figure~\ref{fig:UPD}(c)) at a potential of 0.46~V SHE.
Experimentally, the two voltammogram peaks are at approximately $(0.63\pm 0.04)$
and $(0.51\pm 0.02)$~V SHE, averaging over the forward and reverse direction sweeps.
Therefore chlorine desorption and full-copper-monolayer formation are plausible explanations\cite{ CuPtCl-FullMonolayer-Xrays,CuPtCl-FullMonolayer-LEED1, CuPtCl-FullMonolayer-LEED2}
of the two peaks, with our first-principles predictions reproducing well
the peak spacing (0.09 eV versus 0.12 eV in experiment), and placing the
absolute locations of the peaks to within 0.07 eV.
Similar accuracy has been achieved in comparison to experiment
for onset potentials and product selectivity in CO reduction on
Cu(111)\cite{COreductionCu111,CO-Cu111-selectivity} and the oxygen
evolution reaction on IrO2(110)\cite{IrO2-OER} in concurrent work using exactly
the same calculation protocol as here: fixed-potential DFT in JDFTx using
the PBE exchange-correlation functional and the CANDLE solvation model.
The partial 2x2 monolayer of copper (Figure~\ref{fig:UPD}(d)) proposed by
others\cite{CuPtCl-PartialMonolayer-LEEDandXray} as the reason for the second peak
is not the most stable configuration in our calculations at any potential,
lying a significant 0.3~eV above the other phases at relevant potentials.

For comparison, Figure~\ref{fig:UPD}(b) shows the analogous results that would
be obtained using only conventional vacuum calculations. In the above formalism,
this corresponds to assuming $\Phi^\alpha(\mu) \approx A^\alpha - \mu N_e^\alpha$,
where $A^\alpha$ is the Helmholtz energy from a neutral vacuum DFT calculation
of configuration $\alpha$ containing $N_e^\alpha$ valence electrons.
This approximation results in a single transition directly from the Cl-covered Pt surface to the
one with a copper monolayer, predicting a single voltammogram peak in disagreement with experiment.
Accurate predictions for electrochemical systems therefore require treating
charged configurations stabilized by the electrolyte at relevant electron potentials,
now easily accomplished with the methods and algorithms introduced in this work.

\section*{Conclusions}
This work introduces algorithms for directly converging DFT calculations
in the grand-canonical ensemble of electrons, where the number of electrons
adjusts to maintain the system at constant electron chemical potential, while
ionic response in a continuum solvation model of electrolyte keeps the system neutral .
We show that, with appropriate modifications, grand canonical versions
of both the self-consistent field (GC-SCF) method as well as
direct free-energy minimization with auxiliary Hamiltonians (GC-AuxH) method
are able to rapidly converge the grand free energy of electrons.
This substantially improves upon the current state of the art of
running an outer loop over conventional fixed-charge DFT calculations.
With detailed tests of the convergence of all these algorithms,
we show that the GC-AuxH method is the most suitable default choice
exhibiting smooth exponential convergence to the minimum.

Grand-canonical DFT directly mimics the experimental condition in
electrochemical systems, where electrode potential sets the chemical potential
of electrons, and the number of electrons at the electrode surface (including
adsorbates in the electrochemical interface) changes continuously in response.
Describing this change in charge at the surface plays an important role
in accurately modeling several electrochemical phenomena.\cite{FormateOxidation,AnomalousPH,COreductionCu111,UadjustLoop}
Here, we showcase the new algorithms by analyzing the under-potential deposition (UPD)
of copper on platinum in an electrolyte containing chloride ions.
We resolve an old debate about the identity of a second under-potential peak,
showing that partial copper monolayers are not plausible and that
the second peak is due to desorption of chloride ions.
We expect the new methods presented here to substantially advance
the realistic treatment of electrochemical phenomena in first principles calculations.

\begin{acknowledgements}
RS and WAG acknowledge support from the Joint Center for Artificial Photosynthesis (JCAP),
a DOE Energy Innovation Hub, supported through the Office of Science
of the U.S. Department of Energy under Award Number DE-SC0004993.
RS and TAA acknowledge support from the Energy Materials Center at Cornell (EMC$^2$),
an Energy Frontier Research Center funded by the U.S. Department of Energy,
Office of Science, Office of Basic Energy Sciences under Award Number DE-SC0001086.
Calculations in this work used the National Energy Research Scientific Computing Center (NERSC),
a DOE Office of Science User Facility supported by the Office of Science of
the U.S. Department of Energy under Contract No. DE-AC02-05CH11231.
We thank Kendra Letchworth-Weaver, Kathleen Schwarz, Yuan Ping, Hai Xiao,
Tao Cheng, Robert Smith Nielsen and Jason Goodpaster for insightful discussions.
\end{acknowledgements}

\makeatletter{} 
\end{document}